\documentstyle[eqsecnum,aps]{revtex}

\makeatletter
\def\lsim{\mathrel{\mathpalette\gl@align<}}
\def\gsim{\mathrel{\mathpalette\gl@align>}}
\def\gl@align#1#2{\lower.6ex\vbox{\baselineskip\z@skip\lineskip\z@
    \ialign{$\m@th#1\hfil##\hfil$\crcr#2\crcr\sim\crcr}}}

\makeatother
\begin{document}
\draft
\title
{
Transition between Compressible and Incompressible States in \\
Infinite-Layer Fractional Quantum Hall Systems
}
\author{Sei Suzuki\cite{email} and Yoshio Kuramoto}
\address{
Department of Physics, Tohoku University, \\
Sendai 980-8578, Japan
}
\date{\today}
\maketitle
\begin{abstract}
Possible phase transitions between incompressible quantum Hall
states and compressible three-dimensional states are
discussed for infinite-layer electron systems in strong
magnetic field. By variational Monte Carlo calculation,
relative stability of some trial states is studied.
If the inter-layer distance
is large enough, the Laughlin state is stabilized for the
Landau-level filling $\nu=1/3$ of each layer.
When the inter-layer distance is comparable to the
magnetic length, the Laughlin state becomes unstable against
a different fractional quantum Hall state with inter-layer
correlation, and/or against a three-dimensional compressible
state. It is discussed how the quantum phase transition
between them can be controlled in actual systems.
\end{abstract}



\section{Introduction}
Although the fractional quantum Hall effect is
originally a two-dimensional phenomenon
\cite{review1,Laughlin,Tsui,review2},
multi-layer systems have been studied as natural development
\cite{review2,Yoshioka1,Qiu,MacDonald,SongHe}.
Multi-layer systems show some
unusual effects because of new additional degrees of freedom,
namely the inter-layer Coulomb interaction and
the inter-layer electron tunneling.
For example, in the case of the simplest double-layer system,
the "even denominator" fractional
quantum Hall effect with the Landau level filling
$\nu = 1/2 = 1/4 + 1/4$ has been observed \cite{Eisenstein},
and is interpreted in terms of Halperin's $\Psi_{3 3 1}$
state \cite{Yoshioka1}.

One may then naturally ask what happens in infinite
multi-layer systems.
Experimentally,
the integer quantum Hall effect is observed
in a 30-layer GaAs/AlGaAs superlattice
system \cite{Stormer} and also in organic conductors
$\left({\rm TMTSF}\right)_{2}{\rm X}$
\cite{Chamberlin,Cooper,Hannahs,Kang} and
BEDT-TTF salts \cite{Harrison}.
Motivated by these works we study
the ground state of the infinite-layer system under a strong
magnetic field and discuss quantum phase transitions
between incompressible
and compressible states.

We can have the following naive
expectation in advance of quantitative analysis:
In the limit of infinite inter-layer distance $d_{s}$,
layers are completely independent of each other and
the Laughlin state is realized for
appropriate fillings, e.g. $\nu = 1/q$ where $q$ is
an odd integer. However for $d_{s}\sim l_{B}$ where
$l_{B}$ is the magnetic length,
other fractional quantum Hall states
can be realized because of the inter-layer Coulomb
interaction. Furthermore three-dimensional states
are also realizable with increasing
electron tunnelings between layers.
The fractional quantum Hall states
are incompressible, but the last three-dimensional state is
compressible because the Bloch states are formed perpendicular
to layers and the Fermi surface should appear.
Therefore a transition between incompressible and
compressible states is expected at a certain
value of $d_{s}$,
provided that $d_{s}$ is a variable parameter.

Based on the above expectation we introduce three typical
trial states for the ground state with the lowest Landau
level filling $\nu = 1/3$ and compute energies
of them in order to find which one is the ground state.
Under a strong
magnetic field we restrict the single-particle states within
the lowest Landau level, and consider only the Coulomb
energy and tunneling energy. We calculate the
Coulomb energy by the variational Monte Carlo (VMC) method
for the two-dimensional trial
states \cite{MacMillan,Morf}, while
the Hartree-Fock approximation is used for the
three-dimensional trial state.

\section{Model}
We consider a system with the area ${\cal S}$ of each layer
and the height $L_{z}$ as shown in Fig.~\ref{config}.
The total number $N_{z}$ of layers is given by
$N_{z} = L_{z}/d_{s}$.
Then, with the total number $N$ of electrons in the
system, we define the electron
number per layer $N_{\perp}$ by $N_{\perp} = N/N_{z}$.
Taking into account the layer structure of the system,
we assume the positive charge distribution
\begin{equation}
   n_{b}\left({\bf r}\right) = \sum_{l=1}^{N_{z}}
        \frac{N_{\perp}}{{\cal S}}
        \delta\left(z - l d_{s}\right).
\end{equation}
\noindent
This positive charge density corresponds to the uniform
distribution
in each layer and the discrete distribution along the layer.
Under a magnetic field
${\bf B} = B{\bf e}_{z}$ the Landau level
degeneracy $m_{D}$ is given by
$m_{D} = {\cal S}/2\pi l_{B}^{2}$ where
$l_{B} = \left(c\hbar /eB\right)^{1/2}$ is the
magnetic length. The average filling
factor of each layer $\nu$ is defined as $\nu = N_{\perp}/m_{D}$.
Under the strong magnetic field with $\nu\le 1$
we may deal with only the lowest
Landau level (LLL) explicitly and neglect higher levels.
We will discuss possible ground states in the subspace
of the LLL in the following.

Now we write down the Hamiltonian of the system with
$N\to\infty$ as
\begin{equation}
   H = \sum_{i=1}^{N}H_{0 i}
     + \sum_{i<j}\frac{e^{\ast 2}}{|{\bf r}_{i} - {\bf r}_{j}|}
     - \sum_{i=1}^{N}\int d{\bf r}
              \frac{e^{\ast 2}}{|{\bf r}_{i}
              - {\bf r}|}n_{b}\left({\bf r}\right)
     + \frac{1}{2}\int d{\bf r}d{\bf r}'
              \frac{e^{\ast 2}}{|{\bf r} - {\bf r}'|}
              n_{b}\left({\bf r}\right)n_{b}\left({\bf r}'\right).
              \label{Hamiltonian}
\end{equation}
\noindent
Here the one-body part is given by
\begin{equation}
   H_{0 i} =
      \frac{1}{2 m_{\perp}^{\ast}}\left({\bf p}_{\perp i}
         + \frac{e}{c}{\bf A}_{\perp}\left({\bf r}_{i}\right)
                                  \right)^{2}
      + \frac{1}{2 m_{\parallel}^{\ast}} p_{z i}^{2}
      + V_{0}\left(z_{i}\right),
\end{equation}
\noindent
where $m_{\perp}^{\ast}$ and $m_{\parallel}^{\ast}$ are the
effective masses in the layer and along the $z$-direction
respectively. The squared effective charge
$e^{\ast 2}$ is defined by use of the dielectric constant
$\varepsilon$ as $e^{\ast 2} = e^{2}/\varepsilon$.
$V_{0}\left(z\right)$ is the binding potential of electrons to
layers (like a multi square well potential).
We choose two types of configuration as follows:
one is the circular shape and
the other is the rectangular shape.
These shapes are chosen for convenience of calculations.

\subsection{Circular Geometry}
Let us consider a pile of infinitely large layers
(${\cal S} = \infty$) with the circular symmetry and
take the symmetric gauge. For the
($x,y$)-space this geometry is the same as the one
that Laughlin chose \cite{Laughlin}.
We adopt for this geometry the periodic boundary
condition in the $z$-direction. This geometry is convenient for
description of two-dimensional states and also convenient for
the VMC calculation \cite{Morf}.
If we assume that electrons do not transfer
between layers, then the one-particle state is
written as
\begin{equation}
  \psi_{l m}\left({\bf r}\right)
     = \left(2^{m+1} \pi m! l_{B}^{2m+2}\right)^{-1/2}
       \xi^{m}\exp\left[-\frac{|\xi|^{2}}{4l_{B}^{2}}\right]
       \phi_{0}\left(z-ld_{s}\right),
\end{equation}
\noindent
where $\xi = x - iy$ is the
complex coordinate in the ($x,y$)-space and
$\phi_{0}\left(z - ld_{s}\right)$ defined by
(\ref{phizero}) is the normalized wave function
localized at the layer $l$.
Since the uniform positive charge
is inconvenient for finite particles which we actually
study numerically, we take an alternative positive charge
density $\tilde{n}_{b}\left({\bf r}\right)$ which has
the same distribution as that of electrons.
Namely with
$P\left(\left\{{\bf r}_{i}\right\}\right) =
|\psi \left(\left\{{\bf r}_{i}\right\}\right)|^{2}$
being the probability distribution function of electrons,
we require
\begin{equation}
   \tilde{n}_{b}\left({\bf r}\right) = \sum_{j=1}^{N}
          \int\left(\prod_{i}d{\bf r}_{i}\right)
          \delta\left({\bf r} - {\bf r}_{j}\right)
          P\left(\left\{{\bf r}_{i}\right\}\right).
\end{equation}
This positive charge distribution tends to the uniform one
in the infinite-size
limit.

We now discuss the Coulomb potential with periodic
boundary condition along the $z$-axis. Taking account of mirror
images we obtain the potential as
\begin{equation}
   \sum_{n=-\infty}^{\infty}\frac{e^{\ast 2}}
       {\left[\left({\bf r}_{\perp} - {\bf r}_{\perp}'\right)^{2}
              + \left(z - z' + nL_{z}\right)^{2}
        \right]^{1/2}}.
\end{equation}
\noindent
The summation over $n$ is divergent logarithmically.
This divergence should be canceled by the attractive
part in the final result. In this paper
we replace it by the following one:
\begin{equation}
   {\cal V}\left({\bf r} - {\bf r}'\right) =
      \frac{e^{\ast 2}}
      {\left[\left({\bf r}_{\perp} - {\bf r}_{\perp}'\right)^{2}
             + \left(\frac{L_{z}}{\pi}
               \sin\left(\frac{z-z'}{L_{z}}\pi\right)\right)^{2}
       \right]^{1/2}}.
\end{equation}
\noindent
This potential has the right periodicity along the $z$-axis
and becomes
equivalent to the Coulomb potential in the limit of
$L_{z}\to\infty$. The Hamiltonian with finite $N$ is represented
as follows:
\begin{equation}
   H = \sum_{i=1}^{N}H_{0 i}
           + \sum_{i<j}{\cal V}
             \left({\bf r}_{i} - {\bf r}_{j}\right)
           - \sum_{i=1}^{N}\int d{\bf r}
             {\cal V}\left({\bf r}_{i} - {\bf r}\right)
             \tilde{n}_{b}\left({\bf r}\right)
           + \frac{1}{2}\int d{\bf r}d{\bf r}'
             {\cal V}\left({\bf r} - {\bf r}'\right)
             \tilde{n}_{b}\left({\bf r}\right)
             \tilde{n}_{b}\left({\bf r}'\right).
           \label{Hamiltonian1}
\end{equation}
\noindent
The expectation value is given by
\begin{eqnarray}
   \langle H\rangle &= &N\Omega
      + \sum_{i<j}
      \int\left(\prod_{i=1}^{N}d{\bf r}_{i}\right)
      {\cal V}\left({\bf r}_{i} - {\bf r}_{j}\right)
      P\left(\left\{{\bf r}_{i}\right\}\right)
      \nonumber \\
                    &~ &- \frac{1}{2}\sum_{i,j}
      \int\left(\prod_{i=1}^{N}d{\bf r}_{i}
      d{\bf r}_{i}'\right)
      {\cal V}\left({\bf r}_{i} - {\bf r}_{j}'\right)
      P\left(\left\{{\bf r}_{i}\right\}\right)
      P\left(\left\{{\bf r}_{i}'\right\}\right),
\end{eqnarray}
\noindent
where $\Omega$ is a constant representing the single-particle
energy. It is given by
\begin{equation}
   \Omega = \frac{1}{2}\omega_{c}
      + \int d z
      \phi^{\ast}_{0}\left(z\right)
      \left(\frac{p_{z}^{2}}{2 m_{\parallel}^{\ast}}
            + V_{0}\left(z\right)\right)
      \phi_{0}\left(z\right),
\end{equation}
where we have assumed that the tunneling along the $z$-direction is
suppressed.

\subsection{Rectangular Geometry}
Next we consider a pile of rectangular
(${\cal S} = L_{x}\times L_{y}$)
layers. We impose the periodic boundary condition in each
($x,y$ and $z$) direction and choose the Landau gauge.
We obtain $m_{D} = L_{x}L_{y}/2\pi l_{B}^{2}$. In the case
of three-dimensional states, this geometry is
convenient. When there are tunnelings,
the one electron state is written as
\begin{equation}
   \phi_{j k}\left({\bf r}\right)
      = \phi_{\perp j}\left({\bf r}_{\perp}\right)
        \phi_{\parallel k}\left(z\right),
\end{equation}
\noindent
where $j$
($= 1,2,\cdots,m_{D}$) indicates the degree of freedom in the
LLL, and $k = 2\pi n/L_{z}$
($n = -N_{z}/2,-N_{z}/2+1,\cdots,N_{z}/2-1$) is the wave
number along the $z$-direction. The
$({\bf r}_{\perp})$-dependent part
$\phi_{\perp j}\left({\bf r}_{\perp}\right)$
is the Landau wave function \cite{Yoshioka2}:
\begin{equation}
   \phi_{\perp j} \left( {\bf r}_{\perp} \right)
      = \left(\frac{1}{L_{y} l_{B} \sqrt{\pi}}\right)^{1/2}
        \sum_{n=-\infty}^{\infty}\exp
        \left[ - i \kappa_{j+m_Dn} y
            - \frac{\left(x - X_{j+m_{D}n}\right)^2}{2 l_{B}^{2}}
        \right], \label{Landauwf}
\end{equation}
\noindent
where $\kappa_{j}=2\pi j/L_{y}$ and
$X_{j}=l_{B}^{2}\kappa_{j}$. For the $z$-dependent part
$\phi_{\parallel k}\left(z\right)$
we make the Bloch function by use of $\phi_{0}$ as
\begin{equation}
   \phi_{\parallel k} \left(z\right) = \frac{1}{\sqrt{N_z}}
      \sum_{n_z = -\infty}^{\infty}
      \sum_{n = 1}^{N_{z}} e^{i k n d_{s}}
      \phi_{0} \left( z - n d_{s} - n_z L_z \right),
\end{equation}
\noindent
where we take the following localized
function:
\begin{equation}
   \phi_{0} \left(z\right)
       = \left(\frac{1}{{\epsilon}^{2} d_{s}^{2} \pi} \right)^{1/4}
       e^{-z^{2}/2{\epsilon}^{2}d_{s}^{2}}, \label{phizero}
\end{equation}
\noindent
with $\epsilon$ being positive infinitesimal.
Under the tight-binding approximation
$\phi_{j k}\left({\bf r}\right)$ becomes the
eigenstate of the one-particle Hamiltonian $H_{0 i}$.
We introduce the creation and annihilation 
operators, $a_{j k}^{\dagger}$ and $a_{j k}$, corresponding to
$\phi_{j k}\left({\bf r}\right)$.
Then the total Hamiltonian of
this system is given in second quantized form by
\begin{eqnarray}
   {\cal H} &= &\sum_{j}\sum_{k}
       \left(E_{k} - \frac{N}{2V}\frac{e^{\ast 2} d_{s}^{2}}{\pi}
       2\zeta\left(2\right)\right)
       a_{j k}^{\dagger}a_{j k} \nonumber \\
           &~ &+ \frac{1}{2V}\sum_{{\bf q}\neq 0}
       v\left(q\right)
       \left[\rho\left({\bf q}\right)\rho\left(-{\bf q}\right)
             - \rho\left(0\right)
               \exp\left[- \frac{l_{B}^{2}q_{\perp}^{2}}{2}
                         - \frac{\epsilon^{2}d_{s}^{2}q_{z}^{2}}{2}
                   \right]
       \right]. \label{Hamiltonian2}
\end{eqnarray}
\noindent
Here $E_{k} = \Omega - 2t\cos k d_{s}$ is the eigenvalue of the
one-electron state with the inter-layer transfer $-t$,
$\zeta\left(n\right)$ is the Riemann's zeta function,
$v\left(q\right) = 4\pi e^{\ast 2}/q^{2}$ is the
Fourier coefficient of the Coulomb potential,
and $\rho\left({\bf q}\right)$ is the density operator defined by
\begin{equation}
   \rho\left({\bf q}\right) = \int d{\bf r}
      \psi^{\dagger}\left({\bf r}\right)\psi\left({\bf r}\right)
      e^{-i{\bf q}\cdot{\bf r}},
\end{equation}
\noindent
where
$\psi\left({\bf r}\right)
= \sum_{j k}a_{j k}\phi_{j k}\left({\bf r}\right)$
is the electron field operator projected onto the LLL.

\section{Trial States}
We introduce three trial states for the ground
states with $\nu = 1/3$ of the
Hamiltonians (\ref{Hamiltonian1}) and
(\ref{Hamiltonian2}). As mentioned in the previous section,
we take (\ref{Hamiltonian1})
for states without the tunneling (two-dimensional states) and
(\ref{Hamiltonian2}) for states with the tunneling
(three-dimensional states).
Here we remark on the index of position coordinates.
The position coordinates in the Hamiltonian
(\ref{Hamiltonian1}) have the index $i$
($= 1,2,\cdots,N$) of electrons. However when we refer to
two-dimensional states, we make the position coordinates have the
layer index additionally because electrons belonging to different
layers are distinguishable. Hence the
position coordinates have the two indices $l$ and $p$ where
$l$ ($= 0,1,\cdots,N_{z}-1$) is the layer index and
$p$ ($= 1,2,\cdots,N_{\perp}$) is the particle index in a layer.
Indices
$\{i\}$ and $\{l,p\}$ are related to each other by
$i = N_{\perp}l + p$.

The first trial state is a two-dimensional one without the
tunneling represented by
\begin{equation}
   \Psi_{\nu = 1/3}^{L} = \prod_{l=1}^{N_{z}}\left[
         \chi_{\nu = 1/3}^{L}\left(\left\{\xi_{l p}\right\}\right)
         \right]
         \Phi\left(\left\{\left\{z_{l p}\right\}\right\}\right)
         , \label{mltLaughlin}
\end{equation}
\noindent
where $\chi$ and $\Phi$ describe the motion in the
($x,y$)-space and along the
$z$-direction respectively. They are given by
\begin{equation}
   \chi_{\nu = 1/3}^{L}\left(\left\{\xi_{l p}\right\}\right)
      = \prod_{p<p'}
      \left(\xi_{l p} - \xi_{l p'}\right)^{3}
      \exp\left[-\frac{1}{4 l_{B}^{2}}
                 \sum_{p=1}^{N_{\perp}}|\xi_{l p}|^{2}\right],
\end{equation}
\begin{equation}
   \Phi\left(\left\{\left\{z_{l p}\right\}\right\}\right) =
       \prod_{l=1}^{N_{z}}\prod_{p=1}^{N_{\perp}}
       \phi_{0}\left(z_{l p} - l d\right).
\end{equation}
\noindent
Here $(\{\{z_{l p}\}\})$ denotes
$(z_{1 1},\cdots,z_{1 N_{\perp}};\cdots;z_{N_{z} 1},
      \cdots,z_{N_{z} N_{\perp}})$
and $(\{\xi_{l p}\})$ does
$(\xi_{l 1},\xi_{l 2},\cdots,\xi_{l N_{\perp}})$.
We call the state (\ref{mltLaughlin}) the "multi-Laughlin state".
It is the direct product of the Laughlin state with
$\nu = 1/3$ \cite{Laughlin} in each layer
and is expected to be stable
for a large inter-layer distance: $d_{s} \gg l_{B}$.

The second trial state is also a two-dimensional one without
the tunneling. It is represented by
\begin{equation}
   \Psi_{\nu = 1/3}^{\left(1 1 1\right)}
       = \chi_{\nu = 1/3}^{\left(1 1 1\right)}
         \Phi\left(\left\{\left\{z_{l p}\right\}\right\}\right),
\end{equation}
\noindent
where
\begin{equation}
   \chi_{\nu = 1/3}^{\left(1 1 1\right)}
       = \prod_{l=1}^{N_{z}}
           \prod_{p<p'}\left(\xi_{l p} - \xi_{l p'}\right)
           \prod_{p,p'}\left(\xi_{l p} - \xi_{l+1 p'}\right)
           \exp\left[-\frac{1}{4 l_{B}^{2}}
                      \sum_{p=1}^{N_{\perp}}|\xi_{l p}|^{2}
               \right].
\end{equation}
\noindent
We call this state the "(111) state" \cite{Qiu}. This state is
expected to be stable for $d_{s} \sim l_{B}$
where the inter-layer Coulomb interaction becomes
comparable to the intra-layer one.
The numbers (111) indicate the exponents of the Jastrow factor
which controls the strength of the
electron correlation.
The first "1" is the exponent of the Jastrow factor between a
given layer and the nearest lower layer, the second "1" is that
within the layer, and the last "1" is that between the layer
and the nearest upper layer. Generally we can define
($\lambda \mu \lambda$) state for our system.
For example (131)
state is possible for $\nu = 1/5$, and
(212) state is also possible for the same $\nu$.
Physically (212) state should be
unstable because the inter-layer correlation
becomes larger than the intra-layer one in this state.
In the case of $\nu = 1/3$ only the (111) state is allowed
within the subspace of Halperin-type states.

The third state is expected to be stable at
$d\lsim l_{B}$ when electrons transfer between layers. We call
such a state the "itinerant state" and define it
as follows.
With the tunneling, Bloch waves are formed along the
$z$-direction. Then the
simplest candidate for the ground state is the one with the
smallest kinetic (tunneling) energy. Namely the cosine band 
in the $z$-direction is
filled up to the Fermi level~\cite{Abolfath},
with the filling of the Landau level for each wave number
being unity below the Fermi level and zero above it.
We can write it by use of electron
creation operators as
\begin{equation}
   |{\rm itinerant}\rangle = \prod_{j=1}^{m_{D}}
        \prod_{|k|\le k_{F}}
        a_{j k}^{\dagger}|0\rangle, \label{itinerant}
\end{equation}
\noindent
where $k_{F}$
is the Fermi wave number given by
$k_{F} = \nu \pi/d_{s}$. Although we assume $\nu = 1/3$ in
this paper, the itinerant state can be defined for
all $\nu\in [0,1]$.

\section{Ground-State Energies}

\subsection{Multi-Laughlin State and (111) State}
We use the VMC method to calculate
the expectation value of the Coulomb energy numerically
for the multi-Laughlin state
and the (111) state. Technical details are explained in
Appendix. In the case of the multi-Laughlin state,
$\langle H\rangle$ are
reduced to the sum of the two-dimensional result simply
because all layers are independent of each
other, and the inter-layer Coulomb energy
is exactly canceled. We obtain
\begin{eqnarray}
   N \varepsilon_{L} = \langle H\rangle_{L} - N\Omega &= &
       N_{\perp}\sum_{p<p'}\int
       \left(\prod_{p}d{\bf r}_{\perp 1 p}\right)
       \frac{e^{\ast 2}}{|{\bf r}_{\perp 1 p}
                              - {\bf r}_{\perp 1 p'}|}
       |\chi_{\nu = 1/3}^{L}
           \left(\left\{\xi_{1 p}\right\}\right)|^{2}
       \nonumber \\
                                                      &~ &
       - \frac{1}{2}
       N_{\perp}\sum_{p,p'}\int
       \left(\prod_{p}d{\bf r}_{\perp 1 p}\right)
       \frac{e^{\ast 2}}{|{\bf r}_{\perp 1 p}
                              - {\bf r}_{\perp 1 p'}'|}
       |\chi_{\nu = 1/3}^{L}
          \left(\left\{\xi_{1 p}\right\}\right)|^{2}
       |\chi_{\nu = 1/3}^{L}
          \left(\left\{\xi_{1 p}'\right\}\right)|^{2}.
       \label{mLkitaichi}
\end{eqnarray}
On the other hand such cancellations
do not occur for the (111) state,
and the calculation becomes more
complicated because of the inter-layer correlation.
The energy is given by
\begin{eqnarray}
   N\varepsilon_{\left(111\right)} &= &
      \langle H\rangle_{\left(111\right)} - N\Omega \nonumber \\
                                   &= &
      \left(\sum_{l=l'}\sum_{p<p'}
            + \sum_{l<l'}\sum_{p,p'}\right)
      \int\left(\prod_{l}\prod_{p}
            d{\bf r}_{l p}\right)
      {\cal V}\left({\bf r}_{l p} - {\bf r}_{l' p'}\right)
      |\Psi^{\left(111\right)}_{\nu=1/3}
            \left(\left\{\left\{{\bf r}_{lp}
            \right\}\right\}\right)|^{2} \nonumber \\
                                   &~ &
      - \frac{1}{2}
      \sum_{l,l'}\sum_{p,p'}\int\left(\prod_{l}\prod_{p}
            d{\bf r}_{l p}d{\bf r}_{l p}'
            \right)
      {\cal V}\left({\bf r}_{l p} - {\bf r}_{l' p'}'
            \right)
      |\Psi^{\left(111\right)}_{\nu=1/3}(\{\{{\bf r}_{lp}\}\})|^{2}
      |\Psi^{\left(111\right)}_{\nu=1/3}(\{\{{\bf r}_{lp}'\}\})|^{2}.
\end{eqnarray}

Figure~\ref{graph1} shows the
size dependence of the Coulomb energy of the multi-Laughlin state.
The extrapolated value is in good agreement with that in the
literature~\cite{Morf}. The dependence of the Coulomb energy
for the (111) state on the inter-layer distance is
shown in Fig.\ref{graph2}. The calculation was performed for
three cases of finite size systems: $N_{\perp}=20$, $30$ and $42$
with $N_{z}=5$. We find that the energy decreases with decreasing
inter-layer distance. The
dependence on $N_{\perp}$ is small in the region of small
inter-layer distance.

\subsection{Itinerant State}
We calculate the energy of the itinerant state for the Hamiltonian
(\ref{Hamiltonian2}) with use of the Hartree-Fock approximation.
Since we take the homogeneous trial state, the mean field
is given by
\begin{eqnarray}
   \Delta \left( j k ; j' k' \right) &= &
      \langle a^{\dagger}_{j k} a_{j' k'} \rangle
      =
      \langle {\rm itinerant}| a^{\dagger}_{j k}
      a_{j' k'}
      |{\rm itinerant} \rangle \nonumber \\
                                     &= &
      {\delta}_{j j'}
      {\delta}_{k k'} \theta \left( k_F - | k | \right).
      \label{meanfield}
\end{eqnarray}
The mean field for the
interaction part of the Hamiltonian (\ref{Hamiltonian2})
consists of the direct and the exchange interactions.
Of these the energy due to the direct term is written as
\begin{eqnarray}
   \frac{1}{2V}\sum_{{\bf q}\neq 0}v\left(q\right)
      \langle\rho\left({\bf q}\right)\rangle
      \langle\rho\left(-{\bf q}\right)\rangle
      & =& \frac{N^{2}}{2V}\sum_{{\bf q}\neq 0}v\left(q\right)
        \delta_{{\bf q}_{\perp},{\bf g}}\delta_{q_{z},h}
        \nonumber \\
      & =& \frac{N^{2}}{2V}
        \frac{e^{\ast 2}d_{s}^{2}}{\pi}2\zeta\left(2\right),
\end{eqnarray}
\noindent
where we used the relation $\langle\rho\left({\bf q}\right)\rangle =
N\delta_{{\bf q}_{\perp},{\bf g}}\delta_{q_{z},h}$
valid in the
macroscopic limit. Here we have ${\bf g} =
\left(\frac{2\pi}{L_{x}}m_{D}n_{x},
\frac{2\pi}{L_{y}}m_{D}n_{y}\right)$ and
$h = \frac{2\pi}{d_{s}}n_{z}$
with $n_{x}, n_{y}, n_{z}$ integers.
The direct term cancels exactly with the
Coulomb energy of the positive charge
because of homogeneity of the itinerant state within each layer.
Therefore the Hartree-Fock Hamiltonian
is given by
\begin{eqnarray}
   {\cal H}^{HF} &= &\sum_{j}\sum_{k}E_{k}a_{jk}^{\dagger}a_{jk}
        \nonumber \\
                 &~ &- \frac{1}{2V}
        \sum_{j_{1},j_{2}}\sum_{k_{1},k_{2}}
        a_{j_{1}k_{1}}^{\dagger}a_{j_{1}k_{1}}
        \theta\left(k_{F} - |k_{2}|\right)
        \sum_{{\bf q}\neq 0}
        v\left(q\right)
        M\left(j_{1}k_{1},j_{2}k_{2};{\bf q}\right)
        M\left(j_{2}k_{2},j_{1}k_{1};-{\bf q}\right),
\end{eqnarray}
\noindent
where the matrix element for the exchange mean-field is given by
\begin{eqnarray}
   M\left(jk,j'k';{\bf q}\right) &= &\int d{\bf r}
        \Phi_{jk}^{\ast}\left({\bf r}\right)
        \Phi_{j'k'}\left({\bf r}\right)
        e^{i{\bf q}\cdot{\bf r}} \nonumber \\
                                 &= &
        \sum_{h}\sum_{n=-\infty}^{\infty}
        \delta_{k',k-q_{z}+h}
        \delta_{j-j'+m_{D}n,-\frac{L_{y}}{2\pi}q_{y}}
        \exp\left[ - \frac{l_{B}^{2}q_{\perp}^{2}}{4}
                   - \frac{\epsilon^{2}d_{s}^{2}q_{z}^{2}}{4}
                   + iq_{x}X_{j}\right].
\end{eqnarray}
The energy of the itinerant state
is obtained as
\begin{eqnarray}
   E &= &m_{D}\sum_{|k|\le k_{F}}E_{k} \nonumber \\
     &~ &- \frac{m_{D}}{2V}\sum_{{\bf q}\neq 0}
           v\left(q\right)\sum_{k}\sum_{h}
           \theta\left(k_{F}-|k|\right)
           \theta\left(k_{F}-|k-q_{z}+h|\right)
           \exp\left[ - \frac{l_{B}^{2}q_{\perp}^{2}}{2}
                  - \frac{\epsilon^{2}d_{s}^{2}q_{z}^{2}}{2}
                  \right].
\end{eqnarray}
\noindent
The region of summation over $k$ in the above formula is shown
in Fig.~\ref{baaiwake}. In order to derive
the sum $S$ we analyze the cases as follows:
(1) $S = 0$ for $q_{z}-h > 2k_{F}$,
(2) $S = \frac{L_{z}}{2\pi}(k_{F}-(q_{z}-h-k_{F}))$ for
$0 < q_{z}-h \le 2k_{F}$,
(3) $S = \frac{L_{z}}{2\pi}(q_{z}-h+k_{F}-(-k_{F}))$ for
$-2k_{F} < q_{z}-h \le 0$, and
(4) $S = 0$ for $q_{z}-h \le -k_{F}$. As a result we summarize
the four cases into a single expression given by
\begin{equation}
  S=\frac{L_z}{2\pi}2k_F\left(1-\frac{|q_z-h|}{2k_F}\right)
      \theta\left(2k_F-|q_z-h|\right).
\end{equation}
\noindent
Therefore the energy per particle which contains
the kinetic term is given in the thermodynamic limit by
\begin{eqnarray}
   {\varepsilon}^{HF}          &=&
            \alpha - \frac{2 t}{\nu\pi}\sin\left(\nu\pi\right)
            \nonumber \\
                               &~& - \frac{1}{2\left(2\pi\right)^3}
            \sum_{h}\int_{0+}^{\infty}
            2 \pi q_{\perp}d q_{\perp}
            \int_{h-2k_F}^{h+2k_F}d q_z
            v\left(q\right) \left(1-\frac{|q_z-h|}{2k_F}\right)
            \exp\left[-\frac{l_{B}^2}{2}q_{\perp}^2
                     -\frac{{\epsilon}^2d_{s}^2}{2}q_z^2\right]
            \\
                               &=&
            \alpha + \varepsilon_{kin} + \varepsilon_{int}^{HF}.
            \nonumber
\end{eqnarray}
\noindent
At this stage we let $\epsilon\to 0$. It can be shown that
the contribution of large
$q_z$ ($\sim1/\epsilon d_{s}$) to the integration is negligible.
To see this we observe that
$v\left(q\right) = 1/\left(q_{\perp}^2+q_z^2\right)$
becomes ${\cal O}\left(\epsilon^{2}\right)$ and
summation over $h$ converges in the limit of
$\epsilon\to 0$.
Therefore, $\varepsilon_{int}^{HF}$ becomes
\begin{equation}
   {\varepsilon}_{int}^{HF} = -\frac{1}{2\left(2\pi\right)^2}
     \int_{0+}^{\infty}
     q_{\perp}dq_{\perp}\int_{-2k_F}^{2k_F}dq_z
     \sum_{h}\frac{4\pi e^{\ast 2}}
     {q_{\perp}^2+\left(q_z+h\right)^2}
     \left(1-\frac{|q_z|}{2k_F}\right)
     \exp\left(-\frac{l_{B}^2}{2}q_{\perp}^2\right).
\end{equation}
\noindent
With use of the following formula:
\begin{equation}
   \sum_{n=-\infty}^{\infty}\frac{1}{x^2+\left(y+2\pi n\right)^2} =
    \frac{\sinh x}{2x\left(\cosh x-\cos y\right)},
\end{equation}
\noindent
we can accomplish the summation over $h$ as
\begin{equation}
   \sum_{h}\frac{1}{q_{\perp}^2+\left(q_z+h\right)^2} =
   \frac{d_{s}^2\sinh d_{s}q_{\perp}}{2d_{s}q_{\perp}
   \left(\cosh d_{s}q_{\perp}-\cos d_{s}q_z\right)}.
\end{equation}
\noindent
Consequently we obtain the following result:
\begin{equation}
   {\varepsilon}_{int}^{HF} = -\frac{e^{\ast 2}}{l_{B}}
       \frac{1}{2\pi}
       \frac{1}{\bar{d}_{s}}
       \int_{0}^{\infty} d s \int_{0}^{2\pi/3} d t
       \left( 1-\frac{t}{2\pi\nu}\right)
       \frac{\sinh s}{\cosh s-\cos t}
       \exp\left( - \frac{s^2}{2\bar{d}_{s}^2}\right).
\end{equation}
\noindent
where $\bar{d}_{s} = d_{s}/l_{B}$ denotes the inter-layer distance
in units
of the magnetic length.
Figure~\ref{graph3} shows $\varepsilon_{int}^{HF}$ as a function of
the inter-layer distance. We remark that
$\varepsilon_{int}^{HF}$ decreases
with decreasing $d_{s}$. This result originates from
the fact that the inter-layer Coulomb interaction becomes more
effective as layers get closer.

\subsection{Comparison of Ground-State Energies}
We first compare the Coulomb energy per particle of
the three states.
Figure~\ref{graph4} shows the results. The energy of the (111)
state is obtained by the VMC
calculation for 5 layers with 42 particles
in each layer. For the multi-Laughlin and the itinerant
states the result is shown
for the infinite-size system. We remark
the energy of the multi-Laughlin state does not depend on
the inter-layer distance since all layers are independent of
each other and electrons feel other layers neutral.
In the region of $d_{s}\gsim l_{B}$ the Coulomb energy of the
multi-Laughlin state is the lowest.
This is evident because the system becomes a pile of
almost independent two-dimensional layers in such a region.
In the region of $d_{s}\lsim l_{B}$
the (111) state has the lowest Coulomb energy.
This result originates from the fact that the
(111) state gains the inter-layer Coulomb energy at the
cost of the intra-layer one in the region
where intra- and inter-layer interactions are
comparable. With respect to the itinerant state, the Coulomb
energy decreases in the region of
$d_{s}\lsim l_{B}$. There is however no region where it is the
lowest.

So far we have considered only the Coulomb energy.
We must take into account the tunneling energy of
the itinerant state in order to compare the total energies of the
three states. Let us assume that the transfer energy $t$ depends
on the inter-layer distance as
\begin{equation}
   t = A\exp\left( - \alpha d_{s}\right), \label{tofd}
\end{equation}
\noindent
where $A$ and $\alpha$ are constants. Then
as shown in Fig.~\ref{graph5} we find
a region of $d_{s}$ where
the itinerant state has the lowest total energy.
Here we put appropriate values into $A$ and $\alpha$, namely
$A = 0.2~e^{\ast 2}/l_{B}, \alpha = 1.0~/l_{B}$ for
a case shown as $t_{1}$, and
$A = 0.3~e^{\ast 2}/l_{B}, \alpha = 0.6795~/l_{B}$ for
another case shown as $t_{2}$. The parameters for $t_{2}$ with
$B = 10~T$ are chosen to reproduce the band width of
$4t = 2.5~m{\rm eV} = 0.179~e^{\ast 2}/l_{B}$ at
$d_{s} = 0.226~\AA = 2.8~l_{B}$ which seems appropriate for a
GaAs/AlGaAs superlattice \cite{Stormer}.

On the basis of the comparison we propose a phase diagram as shown
in Fig.~\ref{graph6}
for the ground state of the present system
in the plane of $t$ and
$d_{s}$. If we assume the relation (\ref{tofd}) between
$t$ and $d_{s}$ in the diagram,
possible states follow a curve in
Fig.~\ref{graph6} as $d_{s}$ is varied. Depending on the
parameters $A$ and $\alpha$, there are either single transition
($t_{1}$) or double transitions ($t_{2}$).
Consequently the phase diagram Fig.~\ref{graph6}
is interpreted as follows.
\begin{enumerate}
\item In the region $d_{s}\gg l_{B}$,
      the system is a pile of independent
      two-dimensional layers and the incompressible multi-Laughlin
      state is stable.
\item In the region of $d_{s}\lsim l_{B}$ and the tunneling
      energy is small
      ($t\lsim 0.1~e^{\ast 2}/l_{B}$), the
      (111) state which is also incompressible
      is stabilized instead of the multi-Laughlin state
      by the inter-layer correlation.
\item When the tunneling energy is large,
      the compressible itinerant state is stabilized by
      competition of the tunneling energy and
      the Coulomb energy.
\end{enumerate}

\section{Discussion}
We compared the result of the (111) state for a finite size
system and that of the multi-Laughlin and the itinerant
states for the infinite size system. The finite size
($N_{\perp}$ and $N_{z}$) correction should be taken into
account in the (111) state for more quantitative discussion.
While we have no information with respect to the
dependence on $N_{z}$,
the following remark is given from Fig.~\ref{graph2} about
the dependence on $N_{\perp}$. When one increases
$N_{\perp}$ the Coulomb energy much
increases in the region of $d_{s}\gg l_{B}$. While for
$d_{s}\sim l_{B}$ the Coulomb energy looks almost
independent of $N_{\perp}$ within the error bars.
Since the boundaries of the (111) state occurs with
$d_{s}/l_{B}<2$ in Fig.~\ref{graph6}, the finite-size correction
should not change the results qualitatively.

It should be possible to observe transitions in actual systems by
adding a pressure to the system or tilting the magnetic
field. Since adding a pressure corresponds to control of
$d_{s}$, a change of parameters along a curve like
$t_{1}$ or $t_{2}$
in the diagram Fig.~\ref{graph6} is expected.
On the other hand tilting the magnetic field
corresponds to control of $t$
independent of $d_{s}$ because the horizontal component of
the magnetic field works on
electrons to localize in the vertical
direction and reduces $t$. Then the parameter change along a
vertical straight line at a certain $d_{s}$ in Fig.~\ref{graph6}
is expected.

The transition between incompressible and
compressible states is analogous to the Mott transition in
the Hubbard model.
Namely the incompressible multi-Laughlin and (111) states
correspond to an insulator, while the compressible itinerant state
corresponds to a metal. The transition between incompressible
and compressible states occurs due to competition of the
transfer
(tunneling) and the on site (intra-layer) Coulomb energy. In
our model we estimate that the critical transfer energy is
almost $0.1~e^{\ast 2}/l_{B}$.

Although we discussed only three trial states for
$\nu=1/3$ in the present study,
we cannot exclude the possibility of other states.
In terms of two-dimensional incompressible states
for $\nu=1/3$, it is difficult to construct
states more stable than the multi-Laughlin and
the (111) states within the simple family of Jastrow-type
functions. One may however ask whether other states with
three-body correlation or even more correlations
are more favorable energetically. For
three-dimensional compressible states,
the itinerant state is taken to be
the simplest homogeneous state.
Then inhomogeneous states like CDW \cite{Kuramoto,Halperin}
should also be
discussed for more refined argument.
Furthermore we have neglected the spin degree of
freedom which may be relevant to some systems. For example
the SDW state, which accounts for the QHE of
$\left({\rm TMTSF}\right)_{2}{\rm X}$ \cite{Kang},
can be a relevant state depending on parameters in the system.

\section*{Acknowledgments}
One of the authors (S. S) would like to thank H. Yokoyama 
and S. Tokizaki for instruction of the VMC method and useful 
discussion, and M. Arikawa for pertinent suggestions. 

\appendix
\section*{VMC Method for the (111) State}
We describe the method of numerical calculation for the
(111) state in this appendix.
The expectation value of the Coulomb energy for the (111) state
is represented by
\begin{eqnarray}
   N\varepsilon_{\left(111\right)} &= &\int
     \left(\prod_{l}\prod_{p}d{\bf r}_{l p}\right)
     \left(\sum_{l=l'}\sum_{p<p'} + \sum_{l<l'}\sum_{p,p'}\right)
     {\cal V}\left({\bf r}_{l p} - {\bf r}_{l' p'}\right)
     |\Psi_{\nu = 1/3}^{\left(111\right)}
       \left(\left\{\left\{{\bf r}_{l p}
           \right\}\right\}\right)|^{2}
     \nonumber \\
                                  &~ &-\frac{1}{2}\int
     \left(\prod_{l}\prod_{p}d{\bf r}_{l p}d{\bf r}_{l p}'\right)
     \sum_{l,l'}\sum_{p,p'}
     {\cal V}\left({\bf r}_{l p} - {\bf r}_{l' p'}'\right)
     |\Psi_{\nu = 1/3}^{\left(111\right)}
       \left(\left\{\left\{{\bf r}_{l p}\right\}\right\}\right)|^{2}
     |\Psi_{\nu = 1/3}^{\left(111\right)}
       \left(\{\{{\bf r}_{l p}'\}\}\right)|^{2} \nonumber \\
                                  &= &V_{1} + V_{2}.
     \label{11HCoulomb}
\end{eqnarray}
\noindent
We can obtain the numerical value of
$\varepsilon_{\left(111\right)}$ for finite size
(finite particles) systems by the VMC method.
In evaluating the second term with minimum statistical
errors, we separate the
variables in it and reduce the expectation value of the
quantity between two bodies to that of one body.
For this purpose we introduce the
Fourier transform of the
Coulomb potential $\tilde{\cal V}$, one
body distribution function $f$, and its Fourier transform
$g$ as follows:
\begin{equation}
   {\cal V}\left({\bf r} - {\bf r}'\right) =
     \frac{1}{\left(2\pi\right)^{2}}\int d{\bf q}_{\perp}
     \tilde{{\cal V}}_{\bar{z} \bar{z}'}\left(q_{\perp}\right)
     e^{i{\bf q}_{\perp}\cdot
          \left({\bf r}_{\perp} - {\bf r}_{\perp}'\right)},
\end{equation}
\begin{equation}
   \tilde{{\cal V}}_{\bar{z} \bar{z}'}\left(q_{\perp}\right) =
       \frac{4\pi e^{\ast 2}}{q_{\perp}}
       \exp\left[-q_{\perp}\frac{L_{z}}{\pi}
               \sin\left(\frac{|\bar{z}-\bar{z}'|}{N_{z}}\pi\right)\right],
\end{equation}
\begin{equation}
   f\left(r_{\perp}\right) = \int
       \left(\prod_{l}\prod_{p}d{\bf r}_{l p}\right)
       \delta\left({\bf r}_{\perp l p} - {\bf r}_{\perp}\right)
       |\Psi_{\nu = 1/3}^{\left(111\right)}|^{2},
\end{equation}
\begin{eqnarray}
   g\left(q_{\perp}\right) &= &\int d{\bf r}_{\perp}
       f\left(r_{\perp}\right)
       e^{ - i{\bf q}_{\perp}\cdot{\bf r}_{\perp}} \nonumber \\
                           &= &\int
       \left(\prod_{l}\prod_{p}d{\bf r}_{l p}\right)
       e^{ - i{\bf q}_{\perp}\cdot{\bf r}_{\perp l p}}
       |\Psi_{\nu = 1/3}^{\left(111\right)}|^{2},
\end{eqnarray}
\noindent
where we wrote $z/d_{s}$ as $\bar{z}$.
Then the second term $V_{2}$ in (\ref{11HCoulomb}) is arranged
into
\begin{eqnarray}
   V_{2} &= &- \frac{1}{2}
       \frac{1}{\left(2\pi\right)^{2}}\int_{0}^{\infty}
       N_{\perp}^{2}\sum_{l,l'}
       2\pi q_{\perp}dq_{\perp}
       \tilde{{\cal V}}_{l l'}\left(q_{\perp}\right)
       g\left(q_{\perp}\right)^{2} \nonumber \\
         &= &- \frac{1}{2}
       \frac{N N_{\perp}}{\left(2\pi\right)^{2}}\sum_{l=1}^{N_{z}}
       \int_{0}^{\infty} 2\pi q_{\perp}dq_{\perp}
       \tilde{{\cal V}}_{1 l}\left(q_{\perp}\right)
       g\left(q_{\perp}\right)^{2}.
\end{eqnarray}
\noindent
We use this formula to obtain the numerical value of $V_{2}$.
Figure~\ref{gq13} shows an example of numerical results.
The integral with respect to $q_{\perp}$ is performed
by the trapezoidal rule because only discrete values of
$g\left(q_{\perp}\right)$ are obtained by the VMC calculation.
We cut off the upper region in the integral at a
finite value $q_{c}~(\gg 1/l_{B})$ because the integrand
converges to zero near $q_{\perp}\sim q_{c}$
and its contribution to the integral is negligible in the
region $q_{c}<q_{\perp}<\infty$.


\begin{figure}[htbp]
   \caption[configuration]{Configuration of the infinite-layer
           system. Layers are arranged perpendicular
           to the $z$-axis with the inter-layer distance
           $d_{s}$. A magnetic
           field is applied along the $z$-direction.}
   \label{config}
\end{figure}
\begin{figure}
    \caption[graph1]{Coulomb energy per particle of the
                   multi-Laughlin state.
                   The four points correspond to
                   the particle number per layer
                   $N_{\perp} = 20$, $30$, $42$ and $72$.
                   The error bar indicates the Monte Carlo error.
                   We
                   estimate the value in the thermodynamic
                   limit by extrapolating the linear line
                   which fits the four points
                   with the least squares method.
                   The value is $-0.4089\pm0.0002
                   ~e^{\ast 2}/l_{B}$.
                   The energy does
                   not depend on the inter-layer distance
                   in our multi-Laughlin state.}
    \label{graph1}
\end{figure}
\begin{figure}
    \caption[graph2]{Coulomb
                   energy per particle of the (111) state vs. the
                   inter-layer distance.
                   The calculation is done for five layers with
                   $20, 30$ and $42$ particles in each layer.
                   }
    \label{graph2}
\end{figure}
\begin{figure}
  \caption[baaiwake]{Common regions where two step functions,
                    $\theta\left(k_{F} - |k|\right)$ and
                    $\theta\left(k_{F} - |k - q_{z} + g|\right)$,
                    are nonzero.
                    We have common regions in the cases (see text):
                    (2) $0<q_{z}-q\le 2k_{F}$ and
                    (3) $-2k_{F}<q_{z} - g\le 0$. But we do not
                    have a common region
                    in the cases: (1) $q_{z}-g>2k_{F}$ and
                    (4) $q_{z}-g\le -2k_{F}$.}
  \label{baaiwake}
\end{figure}
\begin{figure}
    \caption[graph3]{Coulomb energy
                   per particle of the itinerant
                   state in the thermodynamic limit.
                   }
    \label{graph3}
\end{figure}
\begin{figure}
   \caption[graph4]{Comparison of Coulomb energies $\varepsilon$
           per particle for the three
           states. The result
           on the (111) state is shown for a finite size system
           ($N_{\perp}=42$ and $N_{z}=5$), while
           the other two results
           are shown for the infinite size system
           ($N_{\perp}=\infty$ and $N_{z}=\infty$). The
           (111) state
           has the lowest energy in the region of
           small inter-layer distance
           while the multi-Laughlin state becomes stable for
           large $d_{s}$.
           The crossing of the two states occurs
           near $d_{s}\sim l_{B}$. The itinerant state has no
           region where its Coulomb energy is the lowest.}
   \label{graph4}
\end{figure}
\begin{figure}
    \caption[graph5]{Energy per particle of the itinerant state with
                   the tunneling energy. We assume the dependence of
                   $t$ on the inter-layer
                   distance as
                   $t = A\exp\left(-\alpha d_{s}\right)$ and show
                   two exemplary cases.}
    \label{graph5}
\end{figure}
\begin{figure}
    \caption[graph6]{Proposed phase diagram in the plane of $t$
                   and $d_{s}$.
                   The solid lines are drawn on the basis of
                   the VMC
                   calculation for the (111) state of
                   the finite size system
                   ($N_{\perp}=42$ and $N_{z}=5$). The boundary
                   with the itinerant state is determined from
                   comparison with the Hartree-Fock calculation
                   in the thermodynamic limit.}
    \label{graph6}
\end{figure}
\begin{figure}[htbp]
  \caption[gq]{An example of $g\left(q\right)$ calculated by
          VMC method. Calculation is performed for the system with
          $N_{\perp}=20$ and $N_{z}=5$.}
  \label{gq13}
\end{figure}


\begin{references}
\bibitem{Abolfath}The structure of the band may be modified 
in the presence of the magnetic field as discussed by
M. Abolfath et al.,\ Phys. Rev.\ {\bf B55},10643(1997), 
for the single wide quantum well. This effect enters through 
modification of the transfer integral $t$ in our formulation. 
We do not consider the change explicitly since the effect is 
shown to be small if the magnetic field is perpendicular 
to the layers.
\end{references}
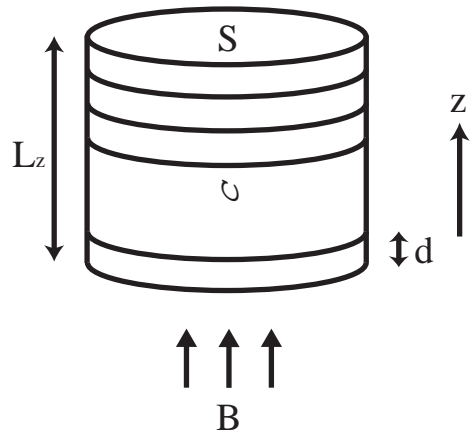
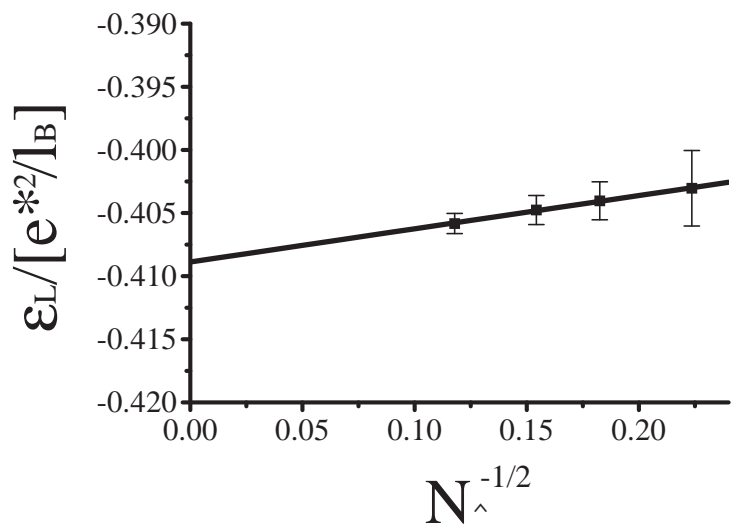
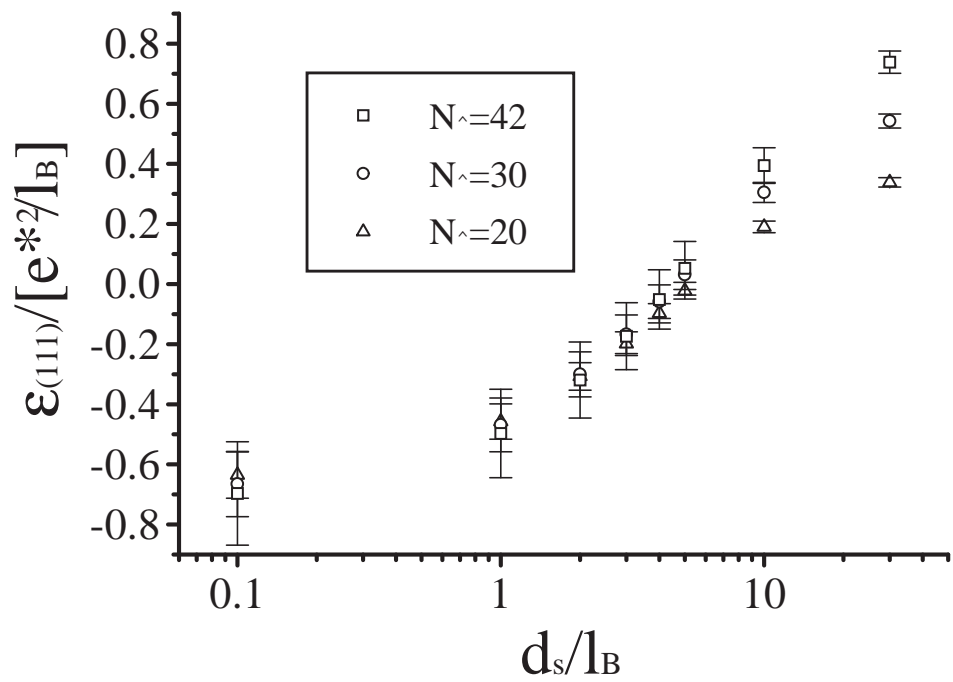
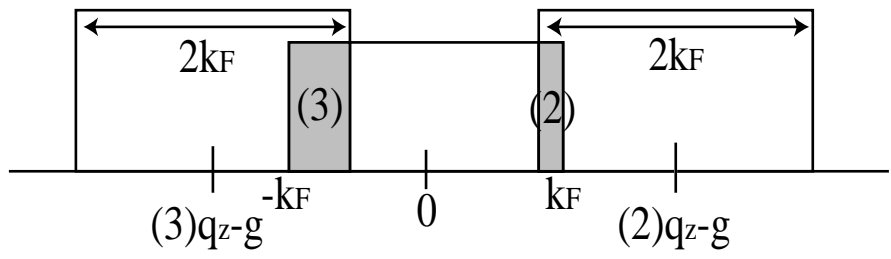
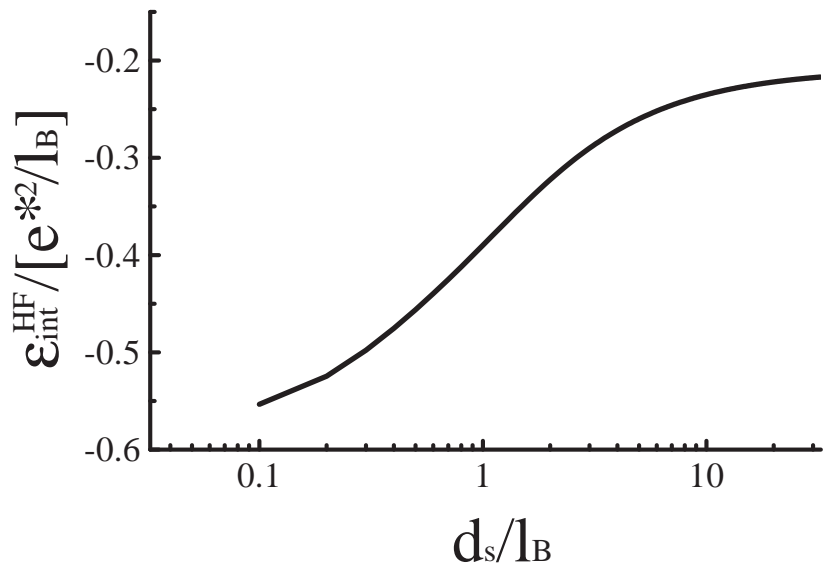
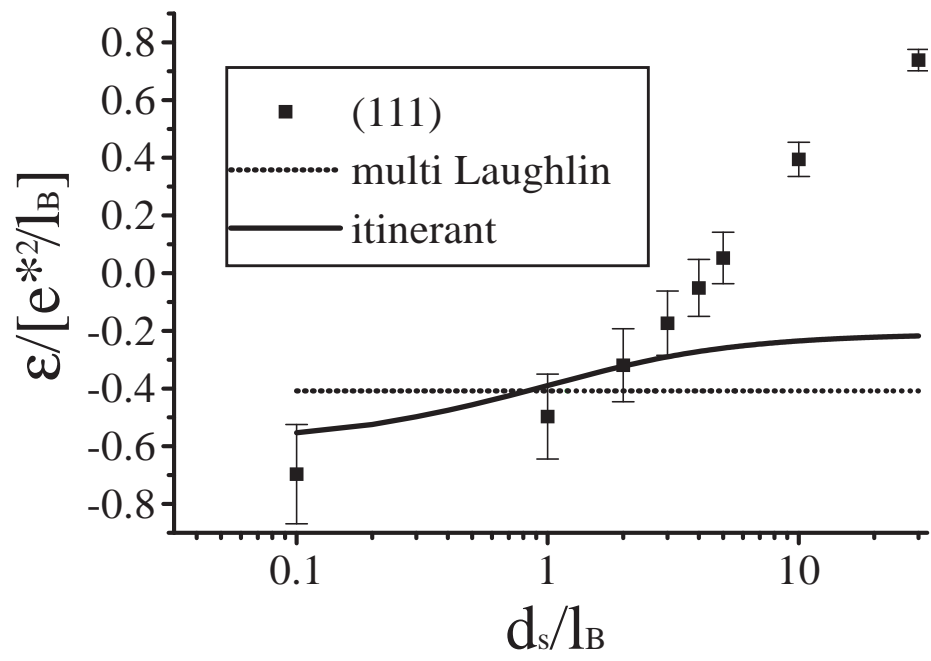
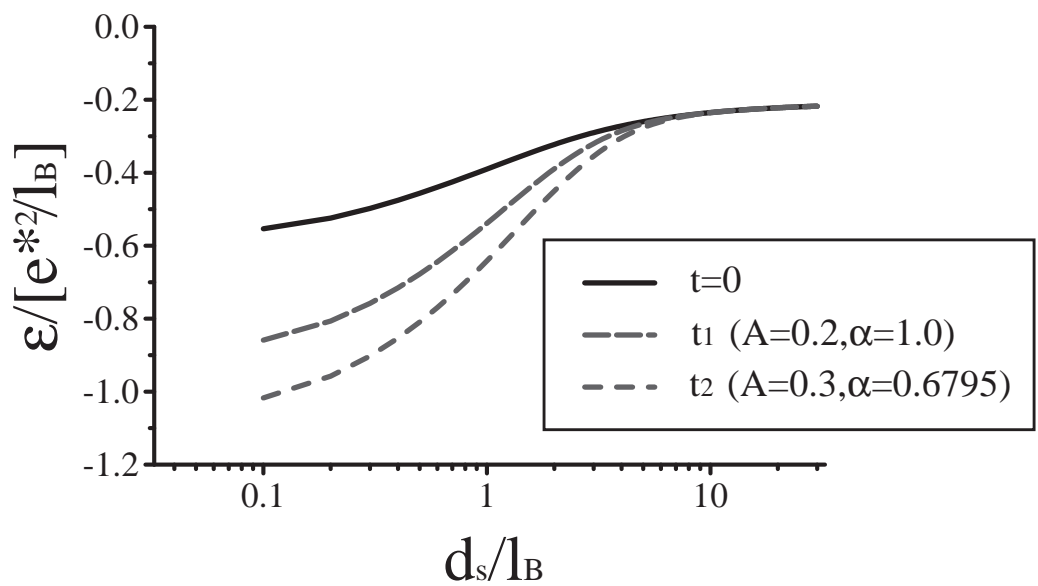
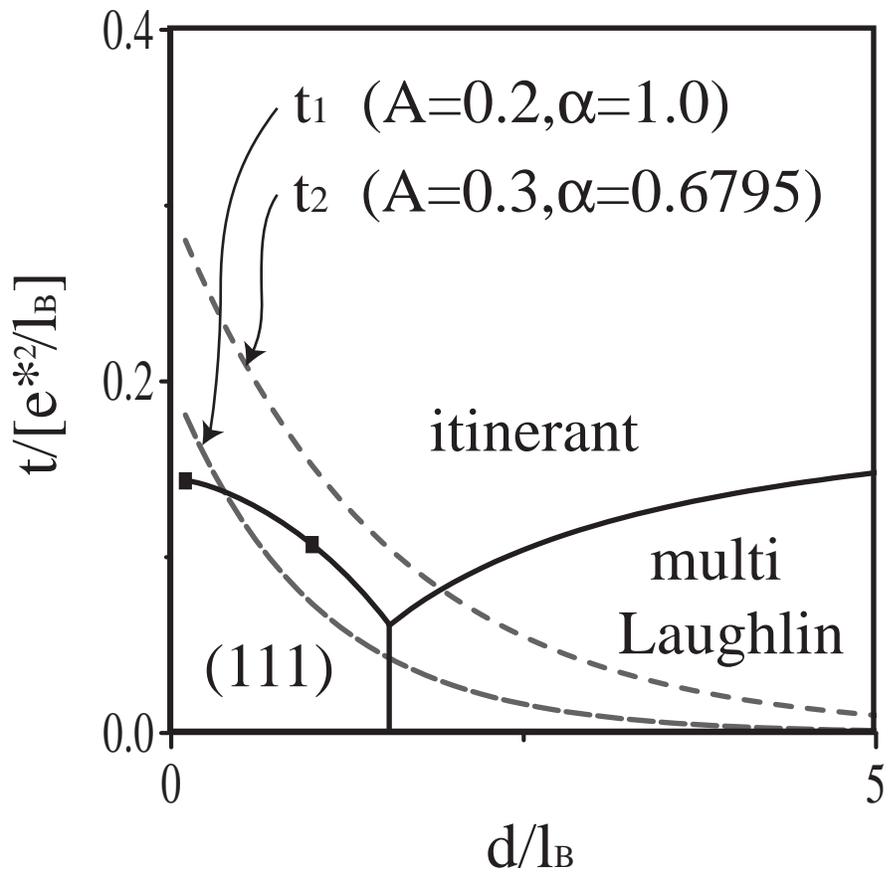
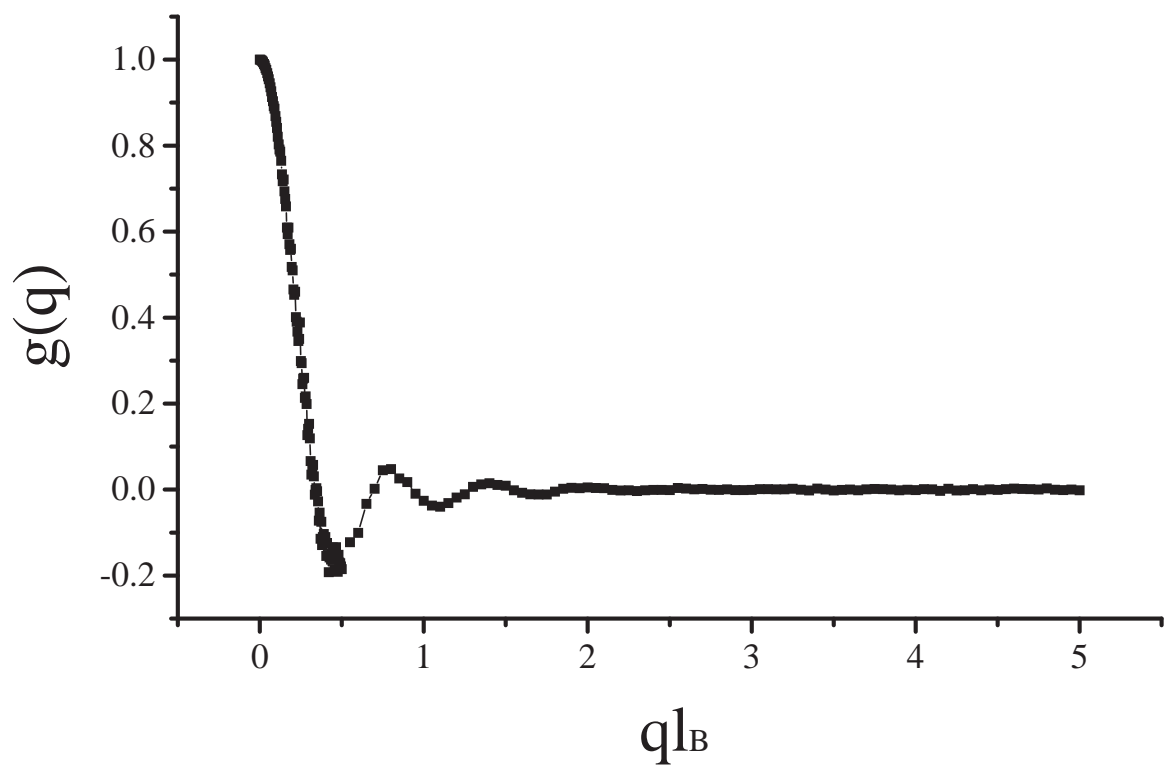
\end{document}